\documentclass[aps,amsmath,amsfonts,11pt]{revtex4}
\usepackage{graphicx}
\usepackage{epstopdf}
\usepackage{epsfig,graphicx}
\usepackage[english]{babel}
\usepackage{amsfonts}
\usepackage{amsmath}
\usepackage{latexsym}
\usepackage{graphics,bm}
\usepackage{dcolumn}
\usepackage{bm}
\usepackage{natbib}
\usepackage{rotating}

\begin{document}

\title{Effect of second-order nonlinearity on quantum coherent oscillations in a quantum dot embedded in a doubly resonant-semiconductor micro-cavity}

\author{ Vijay Bhatt$^{1}$ Pradeep K. Jha$^{1}$ and Aranya B. Bhattacherjee$^{2}$}

\address{$^{1}$  Department of Physics, DDU College, University of Delhi, New Delhi, India }
\address{$^{2}$Department of Physics, Birla Institute of Technology and Science, Pilani, Hyderabad Campus,  Hyderabad - 500078, India}

\begin{abstract}

\end{abstract}

\maketitle

\section{Introduction}
 The recent developments in quantum technology physics have shown tremendous progress in the storage,processing and transfer of quantum information using quantum bits (Qubits) \citep{bou,mon,nie}. Quantum coherence which is a necessary requirement for realistic quantum communication system is extremely fragile and can be destroyed by interaction with the environment.  Semiconductor quantum dots (QDs) embedded in micro-cavity have recently emerged as an attractive candidate for the implementation of quantum computing platforms \citep{pelli,loss,imam,biola,miran}. Instead of the usual two-level real atoms, excitons in the QDs are considered as an alternative two level systems characterized by strong exciton-phonon interactions \citep{hamea,heitz,turck,beso}.
 For practical implementation of quantum information processing based on QDs, it is important to minimize the influence of lattice vibrations which tends to destroy their coherence. Thus it is important to take into account exciton-phonon interactions in the study of quantum-dot cavity system. Experimental observation of vacuum Rabi Oscillations in atomic \citep{brune} as well as in solid state systems \citep{reith,khitro,henn} provides evidence for strong coupling regime in micro cavity systems. Thus QDs embedded in semiconductor micro-cavity have emerged as an exciting platform to study cavity QED \citep{yama,li,wang}. 
         
  Recently proposals have been put forward to use nano structured photonic nanocavities made of $\chi^{(2)}$ nonlinear materials as prospectives  devices for application in quantum information processing, quantum logic gates and all optical switches \citep{arka,Fryett}.One of the main aims of working with such systems is to have a scalable integrated quantum photonic technology with the probability to work at telecommunication wavelengths.  In this paper, we seek to theoretically study the quantum oscillations in a coherently driven quantum dot-cavity system in the presence of a $\chi^{(2)}$ nonlinear substrate and strong exciton-phonon interactions.

\section{Theoretical Model}
	
The system considered here consist of an optical semiconductor microcavity supporting two field modes through nonlinear interaction $g_{nl}$. This nonlinear interaction is provided by the $\chi^{(2)}$ nonlinear substrate introduced into the microcavity as shown in Fig.1. In addition, a quantum dot (QD) is also embedded in the system which interacts with both the optical modes. Due to the nonlinear interaction process, one of the optical mode has two photons at the fundamental frequency while the second mode has a single photon at the second harmonic frequency. In order to ensure phase matching between the two modes, a non-zero spatial overlap between the cavity modes exists \citep{boyd,rivoire}. The semiconductor microcavity considered here can be fabricated using distributed Bragg reflectors(DBR). The light field that is pumped into the cavity is confined in the x-direction by the DBR while the air guiding dielectric provides confinement in the y-z plane \citep{ali,gudat,bhatt}. A $\chi^{(2)}$ nonlinear substrate can be deposited on the GaAs cavity according to known experimental techniques \citep{arka}.

	 \begin{figure}[h]
		\includegraphics [scale=0.8]{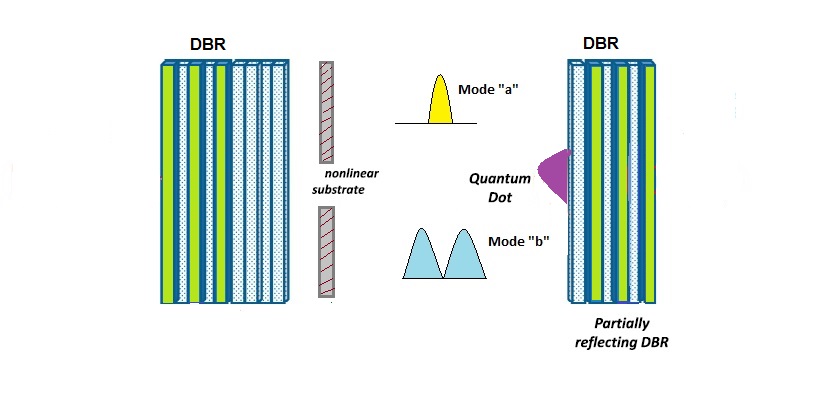}\\
		\caption{ Schematic representation of the setup studied in the text. DBR mirrors is shown in figure and there are two modes of cavity in which  nonlinear substrate is present . The cavity mode "b" interacts with quantum dot. Green and white strips correspond to AlGaAs and GaAs layers respectively}
	\end{figure}

		The embedded semiconductor QD is assumed to be a simple two-level system which consists of the electronic ground state  $ \lvert 1\rangle $ and the lowest-energy electron-hole (exciton) state $\lvert 1\rangle $.
	We consider that this QD via the exciton interacts with both the fundamental mode of frequency $\omega_{a}$ and second harmonic mode of frequency $\omega_{b}= 2 \omega_{a}$. The two level QD can be characterized by the pseudo spin-1/2 operators $\sigma_{\pm}$,$\sigma_{z}$ and the fundamental and second harmonic modes are characterized by the annihilation and creation operators $a (b)$ and $a^{\dagger}$ $(b^{\dagger})$ respectively with Bose commutation relation [$a$,$a^{\dagger}$]=1 ([$b$,$b^{\dagger}$]=1). In this model, we also consider strong coupling of exciton with bulk acoustic phonons. The total Hamiltonian for this coupled photon-exciton-phonon system in the dipole and rotating wave approximation is written as,
	
	 \begin{eqnarray}
	H &=& \hbar\omega_{a}a^{\dagger}a + \hbar\omega_{b}b^{\dagger}b +\hbar\omega_{ex}\sigma_{z}+ \hbar g_{a}(\sigma_{-}a^{\dagger} +\sigma_{+}a) + \hbar g_{b}(\sigma_{-}b^{\dagger} + \sigma_{+}b) \nonumber \\
	&+&\hbar g_{nl}\left(b (a^{\dagger})^2 + b^{\dagger}(a)^2\right) + \hbar\sum_{q}{\omega_q(b_q^{\dagger}b_q)} + \hbar \sigma_{z}\sum_{q}{M_{q}(b_q^{\dagger} + b_{q})}
	\end{eqnarray}
	\vspace*{0.1cm}
	
	The first two terms denote the energy of the fundamental and second harmonic mode respectively. The third term is the exciton energy with $\omega_{ex}$ as the exciton frequency. The fourth and fifth terms are the exciton-fundamental mode photon and exciton - second harmonic photon interaction with coupling constants $g_{a}$ and $g_{b}$ respectively. The sixth term is the nonlinear interaction between the fundamental and second harmonic mode with coupling constant $g_{nl}$ which  can be expressed as 
	
 \begin{equation}
\hbar g_{nl}=\epsilon_{0}\left( \frac{\hbar\omega_{a}}{\epsilon_{0}\epsilon_{r}}\right) ^{3/2} \frac{\chi^{(2)}}{\sqrt{V_{r}}}
\end{equation}
	
	where, $\frac{1}{\sqrt{V_{r}}}$=$\int_{NL}\alpha(r)^{3} dr.$ Note that, the phase matching condition is implicit in the assumption of perfect overlap between the cavity modes. A perfect overlap between the cavity modes ($\alpha_{a}(r)=\alpha_{b}(r)=\alpha(r)$), $\alpha_{a}(r)$ and $\alpha_{b}(r)$ are the normalized field profiles of the cavity modes such that  integration over the whole volume is unity.i.e; $\int\left|\alpha(r)\right|^{2}$ dr =1

	The seventh term denotes the energy of the $q^{th}$ phonon mode having frequency $\omega_{q}$ and creation (annihilation) operator  $b_q^{\dagger}(b_q)$. The last term is the exciton-phonon interaction characterized by the matrix element $M_{q}$.  For the sake of simplicity, when the temperature is low($T < 50K$) \citep{beso,wilson}  we assume that the off-diagonal exciton-phonon interactions are negligible. For an InGaAs quantum dot, the energy separation is about 65MeV from the ground state transition.

     Now first we apply canonical transformation to the Hamiltonian (1) \citep{mahan}

     \vspace*{0.3cm}
    \begin{equation}
    H^{'}= \exp{(S)} H \exp{(-S)},
    \end{equation}

       where the generator is - 
\begin{equation}  
S=\left( S_{z}+\frac{1}{2}\right )\sum_{q}\frac{M_q}{w_q}(b_q^{\dagger} + b_q) .
\end{equation}     

    The transformed Hamiltonian is given by-
    
\begin{equation}
H^{'}=H_{0}^{'}+ H_{I}^{'} ,
\end{equation}

where;
\begin{eqnarray}
H_{0}^{'}&=&\omega_{a}a^{\dagger}a + \omega_{b}b^{\dagger}b + (\omega_{ex}-\Delta)\left(S_{z}+\frac{1}{2}\right)+ \sum_{q}\omega_{q}b_q^{\dagger}b_q ,
\end{eqnarray}

\begin{equation}
H_{I}^{'}=g_a[\sigma_{+}aX^{\dagger}+\sigma{-}a^{\dagger}X] + g_b[\sigma_{+}bX^{\dagger}+\sigma{-}b^{\dagger}X] + g_{nl}[b(a^{\dagger})^2+b^{\dagger}(a)^2] ,
\end{equation}

where 
\begin{equation}
 \Delta=\sum_{q}\frac{M_{q}^2}{w_q} ,
\end{equation}

\begin{equation}
X=exp\left[-\sum_{q}\frac{M_q}{w_q}(b_q^{\dagger}-b_q)\right] ,
\end{equation}

\begin{equation}
X^{\dagger}=X^{-1} .
\end{equation}

    We will now work in the interaction picture with $H_{0}^{'}.$
 The Hamiltonian in the interaction picture is evaluated as,

It is given by,
\begin{equation}
H_{int}=e^{iH_{0}^{'}t}H_{I}^{'}e^{-iH_{0}^{'}t}
\end{equation}
        
Using
\begin{equation}
exp\left[ i\sum_{q}\omega_{q}b_q^{\dagger}b_qt\right] X exp\left[ -i\sum_{q}\omega_{q}b_q^{\dagger}b_qt\right] =exp\left[ -i\sum_{q}\frac{M_q}{w_q}(b_q^{\dagger}e^{i\omega_qt}-b_qe^{-iw_qt})\right] ,
\end{equation}

we have
\begin{eqnarray}
H_{int}&=&g_{a}\left[ \sigma_{+} a X^{\dagger}(t)e^{i\delta_a t} + \sigma_{-} a^{\dagger} X(t)e^{-i\delta_a t}\right] + g_{b}\left[ \sigma_{+} b X^{\dagger}(t)e^{i\delta_b t} + \sigma_{-} b^{\dagger} X(t)e^{-i\delta_b t}\right] \nonumber \\
 &+&g_{nl}\left[ b(a^\dagger)^2 + b^{\dagger}(a)^2\right] 
\end{eqnarray}

	Where;
	\vspace*{0.5cm}
	\hspace*{0.5cm} 		$\delta_a =\omega_{ex}-\omega_{a}-\Delta$ ,
 \hspace*{0.5cm} 	   $\delta_b =\omega_{ex}-2\omega_{a}-\Delta$  

And \hspace*{0.5cm} 	     $X(t) = \exp\left[-\sum_{q}\frac{M_{q}}{\omega_{q}}\left( b_{q}^{\dagger}e^{i \omega_{q}t} - b_{q}e^{-i \omega_{q}t}\right) \right]$

	\vspace*{0.5cm}

		Now we proceed to solve the equation of motion for  $\arrowvert \varPsi(t) \rangle$ ; i.e.
		\vspace*{0.5cm}

	\hspace*{3cm} \vspace*{0.5cm}	$i\frac{d \arrowvert \varPsi(t)\rangle}{dt}= H_{int}\arrowvert \varPsi(t) \rangle$

	  In general , the State vector  $\arrowvert \varPsi(t)\rangle$ is a linear
combination of states  $\arrowvert 1, m, n \rangle$ $\arrowvert ph \rangle$ and   $\arrowvert 2, m, n \rangle$ $\arrowvert ph \rangle$. Here $\arrowvert 2, m, n \rangle$ is the state in which the Quantum dot is in excited state. i.e.  $\arrowvert 1, m, n \rangle$ is ground state. In the excited state Exciton are present.

	 As we are using the Interaction Picture, we use the slowly varying amplitude $C_{1, m ,n ,ph(t)}$ and $C_{2,m,n, ph(t)}$. The State vector is therefore:-
	 
\begin{equation} 	
\arrowvert \varPsi(t)\rangle =\sum_{m,n}\left[ C_{1, m ,n ,ph(t)} \arrowvert 1, m, n \rangle \arrowvert ph \rangle +  C_{2, m ,n ,ph(t)} \arrowvert 2, m, n \rangle \arrowvert ph \rangle\right]
\end{equation}

	( Note that, 'a' operator will act on state n and 'b' operator on  m )

	\vspace*{1cm}

	The Interaction Hamiltonian (13) Can cause transitions between the states as follows-

	\hspace*{2cm}		$\arrowvert 1, m+1, n \rangle  \leftrightarrow \arrowvert 2, m, n \rangle$
	
	\vspace*{0.3cm}
	\hspace*{2cm}		$\arrowvert 1, m, n+1 \rangle   \leftrightarrow \arrowvert 2, m, n \rangle$
	
	\vspace*{0.3cm}
	\hspace*{2cm}	$\arrowvert 1, m, n \rangle    \leftrightarrow   \arrowvert 1, m+2, n-1 \rangle$
	
	\vspace*{0.3cm}
	\hspace*{2cm}	$\arrowvert 1, m-2, n+1 \rangle   \leftrightarrow   \arrowvert 1, m, n \rangle$
	
	\vspace*{0.3cm}
	\hspace*{2cm}       $\arrowvert 2, m, n \rangle \leftrightarrow \arrowvert 2, m+2, n-1 \rangle$
	
	\vspace*{0.3cm}
	\hspace*{2cm}	$\arrowvert 2, m, n \rangle  \leftrightarrow  \arrowvert 2, m-2, n+1 \rangle$

	 \begin{figure}
		\includegraphics [scale=0.8]{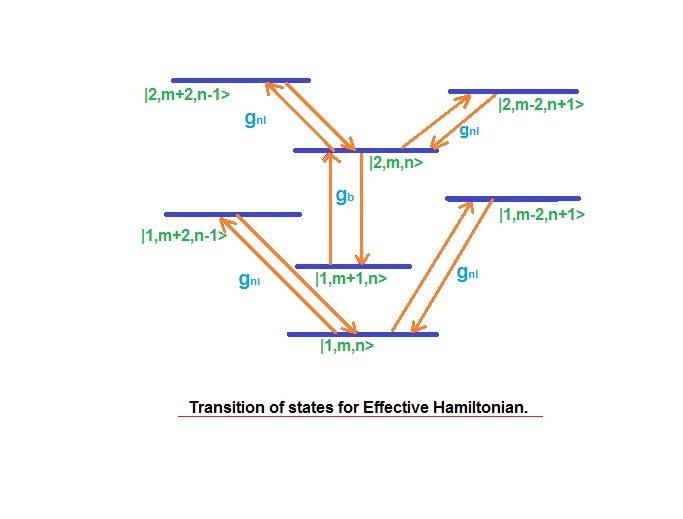}\\
		\caption{ Transition states for effective Hamiltonian.}
	\end{figure}

We can write the Hamiltonian as-

\begin{equation}	
H\arrowvert \varPsi \rangle = H_{a}\arrowvert \varPsi \rangle + H_{b}\arrowvert \varPsi \rangle + H_{nl}\arrowvert \varPsi \rangle
\end{equation}
	
	\vspace*{0.5cm}

		The equations of motion for the Probability Amplitudes are obtained by first
 substituting for $\arrowvert \varPsi(t) \rangle$ and $H_{eff}$ from equation (18) and equation (16) in equation (17),
 	
 	\vspace*{0.4cm}	
 	
 		We obtain the following linear Coupled Equations-	
		
		\vspace*{0.5cm}

\begin{equation}
 i\dot{C}_{2,m+2,n}(t)=g_{nl}C_{2,m,n+1}(t)\sqrt{n+1}\sqrt{m+1}\sqrt{m+2}
\end{equation}

\begin{equation}
i\dot{C}_{2,m,n+1}(t)=g_{nl}C_{2,m+2,n}(t)\sqrt{n+1}\sqrt{m+1}\sqrt{m+2}
\end{equation}

\begin{equation}
i\dot{C}_{1,m,n+1}(t)=g_{b} e^{-\frac{\lambda}{2}} \sqrt{n+1}e^{-i \delta_b t}C_{2,m,n}(t) + g_{nl}C_{1,m+2,n}(t)\sqrt{n+1}\sqrt{m+1}\sqrt{m+2}
\end{equation}

\begin{equation}
i\dot{C}_{2,m,n}(t)=g_{a} e^{-\frac{\lambda}{2}} \sqrt{m+1}e^{i \delta_a t}C_{1,m+1,n}(t) + g_{b}C_{1,m,n+1}(t)\sqrt{n+1}e^{-\frac{\lambda}{2}}e^{i \delta_a t}
\end{equation}

\begin{equation}
i\dot{C}_{1,m+1,n}(t)=g_{a} \sqrt{m+1}e^{-\frac{\lambda}{2}}e^{-i \delta_at}C_{2,m,n}(t)
\end{equation}

\begin{equation}
i\dot{C}_{1,m+2,n}(t)=g_{nl}C_{1,m,n+1}(t)\sqrt{n+1}\sqrt{m+1}\sqrt{m+2}
\end{equation}

	Where,   $\lambda = \sum_{q}( \frac{M_{q}}{\omega_{q}} )^2$ is the Huang-Rhys factor which corresponds to the exciton-phonon interactions. It can be determine by the experiment. \citep{zhu}

		\section{Results}

		The coupled set of equations (20)-(25) above can be solved exactly subject to certain initial conditions. Initially the quantum dot is in the excited state $\arrowvert 2 \rangle$. (i.e, with the presence of the exciton). We have presented here result, after separating these equation into real and imaginary parts (See Appendix-A) and showing the results for Probability amplitude for $C_{2,m,n}$ (i.e excited state)  with respect to time. 

		\vspace*{0.5cm}
			\begin{figure}
			\includegraphics[scale=0.8]{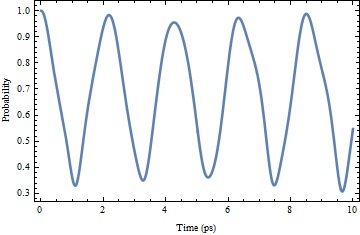}\\
			\caption{The graph between Probability amplitude V/S Time. Parameters are used for graph are- $g_{nl}=2(MeV)$, $\delta_{a}=1$, $\delta_{b}=0.1$, $\lambda=0.01$}
		\end{figure}
	
	In figure-(3,4,5)  we are using parameters $g_{nl}$ , $\delta_{a}$ , $\delta_{b}$ and $\lambda$ , where $g_{nl}$ is nonlinear coupling factor, $\delta_{a}$ is detuning for cavity mode 'a' , $\delta_{b}$ is detuning for cavity mode 'b'. and $\lambda$ is Huang-Rhys factor.

	From figure-(3), we can see that the system is undergoing Quantum Rabi Oscillations. In figure (4), if we change the value of detuning for cavity mode 'a' (i.e $\delta_{a}$) to 0.2  and rest of parameters remain same . We see that Probability of finding the particle in excited state is decreasing by multiple of factor 2. 
	
			\vspace*{0.5cm}
		\begin{figure}
			\includegraphics [scale=0.8]{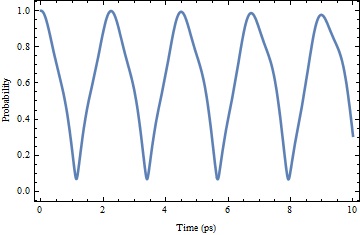}\\
			\caption{parameters used here are $g_{nl}=2(MeV)$, $\delta_a$=0.2, $\delta_b$=0.1, $\lambda$=0.01 }
		\end{figure}

Now for figure (5), when we change the value of Nonlinear coupling factor $g_{nl}$ to 0.5 , We see that a periodic oscillating wave of larger amplitude is generating after two wave of lesser amplitude. It means for a couple of time, the probability of finding the particle in excited state is half and exist between 0.3 and 0.5 in the graph. Time duration is increasing for photons to exist in excited state.

		\begin{figure}
		\includegraphics[scale=0.8]{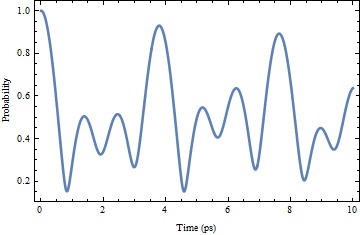}\\
		\caption{The graph between Probability amplitude V/S Time. Parameters are used for graph are- $g_{nl}$=0.5(MeV.), $\delta_{a}$=1, $\delta_{b}$=0.1, $\lambda$=0.01}
	\end{figure}
		
			It should be noted that the couplings of exciton and photons will not become too large so that the Q.Dot cavity system is not in the strong - coupling regime. The value of  Huang-Rhys factor For CdSe quantum dots is $\lambda$ = 1 \citep{turck}, For  InAs/GaAs quantum dots $\lambda$ = 0.015 \citep{heitz} and for other semiconductor quantum dots such as GaAs \citep{arka2} and for InGaAs \citep{huang}, $\lambda$ is even more small.

	\vspace*{1cm}
	
		\section{Conclusion}

		In conclusion, applying Quantum treatment we have discussed the influence of strong exciton-phonon interaction on the quantum Rabi o scillations in a coherently driven quantum dot in a high-Q double -mode cavity in the presence of nonlinear substance. It is found that the coherent oscillations dressed by quantum lattice fluctuations can persist with the coupling constant $g_{a}e^{-\frac{\lambda}{2}}$ and $g_{b}e^{-\frac{\lambda}{2}}$ for two cavity modes. We have investigated that nonlinear substrate affect the Rabi oscillations. When we decrease the nonlinear coupling factor, we see that the number of photons in the higher quantum state exist for a longer time. In this way nonlinear coupling factor increasing the life time of photons to remain in exited state. Our result also indicate that even at the zero temperature, the strong exciton-phonon interactions still affect the quantum coherent oscillation. The interaction of exciton and phonon useful in many applications such as indistinguishable photon generation. It is shown by the result that even at the Zero temperature, the strong exciton- phonon interactions still affect the quantum coherent oscillation significantly and nonlinear coupling factor affecting the oscillation.
		
Thus we proposed a new way of quantum lattice fluctuation on quantum coherence oscillation in the presence of nonlinear coupling factor for Quantum dot-cavity system.
This will make new plateform  for more theoritical and practical applications in QED.

\begin{acknowledgements}
 \textbf{P.K Jha} is thankful to \textbf{SERB-Department of Science and Technology, New Delhi} for the financial support. \textbf{Aranya B. Bhattacherjee} is grateful to \textbf{BITS Pilani, Hyderabad campus} for the facilities to carry out this research.
\end{acknowledgements}

\vspace*{1cm}

	\hspace*{5cm}\textbf{\underline{APPENDIX - A}}\\   

	For separating equation (20)-(25) we let -

	$C_{2,m+2,n}= a_{1} + i a_{2}$ ,  $C_{2,m,n+1}= b_{1} + i b_{2}$ ,  $C_{1,m,n+1}= c_{1} + i c_{2}$ ,   $C_{2,m,n}= d_{1} + i d_{2}$ ,

	$C_{1,m+2,n}= e_{1} + i e_{2}$, $C_{1,m+1,n}= f_{1} + i f_{2}$

	And we get the following equations after separating real and imaginary parts and comparing them.
	\vspace*{0.5cm}
	
	\begin{equation}
	\tag{A1}
	\dot{a_1}= \sqrt{n+1}\sqrt{m+1}\sqrt{m+2}  g_{nl} b_{2},
	\end{equation}
	
	\begin{equation}
	\tag{A2}
	\dot{a_2}= -\sqrt{n+1}\sqrt{m+1}\sqrt{m+2}  g_{nl} b_{1},
	\end{equation}
	
	\begin{equation}
	\tag{A3}
	\dot{b_1}= \sqrt{n+1}\sqrt{m+1}\sqrt{m+2}  g_{nl} a_{2},
	\end{equation}
	
	\begin{equation}
	\tag{A4}
	\dot{b_2}= -\sqrt{n+1}\sqrt{m+1}\sqrt{m+2}  g_{nl} a_{1},
	\end{equation}
	
	\begin{equation}
	\tag{A5}
	\dot{c_1}=\sqrt{n+1}\cos{\delta_{b}t} g_{b} e^{-\lambda/2}d_{2} - \sqrt{n+1}\sin{\delta_{b}t} g_{b} e^{-\lambda/2}d_{1} + g_{nl} A e_{2},
	\end{equation}
	
	\begin{equation}
	\tag{A6}
	\dot{c_2}= -\sqrt{n+1}\cos{\delta_{b}t} g_{b} e^{-\lambda/2}d_{1} - \sqrt{n+1}\sin{\delta_{b}t} g_{b} e^{-\lambda/2}d_{2} - g_{nl} A e_{1},
	\end{equation}

	\begin{align}
	&\;\dot{d_1}= \sqrt{m+1}\cos{\delta_{a}t} g_{a} e^{-\lambda/2}f_{2} + \sqrt{m+1}\sin{\delta_{a}t} g_{a} e^{-\lambda/2}f_{1}\nonumber\\
	+&\;\sqrt{n+1}\cos{\delta_{b}t} g_{b} e^{-\lambda/2}c_{2} + \sqrt{n+1}\sin{\delta_{b}t} g_{b} e^{-\lambda/2}c_{1},\tag{A7}
	\end{align}
	
	\begin{align}
	&\;\dot{d_2}= -\sqrt{m+1}\cos{\delta_{a}t} g_{a} e^{-\lambda/2}f_{1} + \sqrt{m+1}\sin{\delta_{a}t} g_{a} e^{-\lambda/2}f_{2}\nonumber\\
	-&\;\sqrt{n+1}\cos{\delta_{b}t} g_{b} e^{-\lambda/2}c_{1} + \sqrt{n+1}\sin{\delta_{b}t} g_{b} e^{-\lambda/2}c_{2},\tag{A8}
	\end{align}

	\begin{equation}
	\tag{A9}
	\dot{e_1}=\sqrt{n+1}\sqrt{m+1}\sqrt{m+2}  g_{nl} c_{2},
	\end{equation}

	\begin{equation}
	\tag{A10}
	\dot{e_2}= -\sqrt{n+1}\sqrt{m+1}\sqrt{m+2}  g_{nl} c_{1},
	\end{equation}
	
	\begin{equation}
	\tag{A11}
	\dot{f_1}= -\sqrt{m+1}\sin{\delta_{a}t} g_{a} e^{-\lambda/2}d_{1} + \sqrt{m+1}\cos{\delta_{a}t} g_{a} e^{-\lambda/2}d_{2},
	\end{equation}
	
	\begin{equation}
	\tag{A12}
	\dot{f_2}= -\sqrt{m+1}\cos{\delta_{a}t} g_{a} e^{-\lambda/2}d_{1} - \sqrt{m+1}\sin{\delta_{a}t} g_{a} e^{-\lambda/2}d_{2}.
	\end{equation}

	For finding the Probability in excited state   $\arrowvert 2, m, n \rangle$ $\arrowvert ph \rangle$ we calculate $d_{1}^2 + d_{2}^2$.

\end{document}